\documentclass[aps,prl,showpacs,twocolumn]{revtex4}

\usepackage{graphicx}
\usepackage{hyperref}
\usepackage{amsfonts}
\usepackage{amsbsy}
\usepackage{amssymb}
\usepackage{xcolor}
\usepackage[normalem]{ulem}
 \usepackage{float}
 \usepackage{multirow}
 
\begin{document}

%\preprint{}

\title{Entropic extensivity and large deviations in the presence of strong correlations}

\author{Ugur Tirnakli$^{1}$}
\email{ugur.tirnakli@ege.edu.tr}
\author{Mauricio Marques$^{2}$}
\email{mauriciomarx@gmail.com}
\author{Constantino Tsallis$^{2,3,4}$}
\email{tsallis@cbpf.br}

\affiliation{ $^1$Department of Physics, Faculty of Science, Ege University, 35100 Izmir, Turkey \\
$^2$Centro Brasileiro de Pesquisas Fisicas 
and National Institute of Science and Technology for Complex Systems \\
Rua Xavier Sigaud 150, Rio de Janeiro 22290-180, Brazil\\
 $^3$ Santa Fe Institute, 1399 Hyde Park Road, Santa Fe, 
 New Mexico 87501, USA \\
 $^4$ Complexity Science Hub Vienna, Josefst\"adter Strasse 
 39, 1080 Vienna, Austria
 }

\date{\today}

\begin{abstract}
The standard Large Deviation Theory (LDT) mirrors the Boltzmann-Gibbs (BG) factor which 
describes the thermal equilibrium of short-range 
Hamiltonian systems, the velocity distribution of which is Maxwellian. 
It is generically applicable to systems satisfying the 
Central Limit Theorem (CLT), among others. 
When we focus instead on stationary states of typical complex 
systems (e.g., classical long-range Hamiltonian systems), both 
the CLT and LDT need to be generalized. We focus here on a scale-invariant stochastic process 
involving strongly-correlated exchangeable  random variables which, through the Laplace-de Finetti theorem, 
is known to yield a long-tailed $Q$-Gaussian $N\to\infty$ attractor in the space of distributions ($1 < Q<3)$. 
We present strong numerical indications that the corresponding LDT probability distribution is given by 
$P(N,z)=P_0\,e_q^{-r_q(z)N}=P_0[1-(1-q)r_q(z)N]^{1/(1-q)}$ with $q=2-1/Q \in (1,5/3)$. The rate function $r_q(z)$ 
seemingly equals the relative nonadditive $q_r$-entropy per particle, 
with $q_r \simeq \frac{7}{10} + \frac{6}{10}\frac{1}{Q-1}$, thus exhibiting a singularity at $Q=1$ and 
recovering the BG value $q_r=1$ in the $Q \to 3$ limit. Let us emphasize that the extensivity of $r_q(z)N$ 
appears to be verified, consistently with what is expected, from the Legendre structure of thermodynamics, 
for a total entropy.
The present analysis of a relatively simple model somewhat mirroring spin-1/2 long-range-interacting 
ferromagnets (e.g., with strongly anisotropic XY coupling) might be helpful for a deeper understanding of 
nonequilibrium systems with global correlations and other complex systems.
\end{abstract}

%\pacs{05.20.-y,05.10.-a,05.45.-a}

%\keywords{Dynamical systems, Conservative maps, Central limit theorem}
\maketitle

\section{Introduction}
Boltzmann-Gibbs (BG) statistical mechanics yields various 
important relations. Still, it is fair to consider  the Maxwellian distribution of velocities and the 
exponential distribution of energies 
(BG weight or BG factor) as its most important 
fingerprints \cite{statphys1,statphys2}. 
These facts mirror the 
Central Limit Theorem (CLT) \cite{CLT1,CLT2} which leads, 
when the number $N$ of involved random variables increases 
indefinitely, to convergence towards  Gaussian distributions, 
and the Large Deviation Theory (LDT) \cite{LDT1,Peliti,LDT2,Lacomte1,Touchette2009,Lacomte2} which 
characterizes the speed at which Gaussians are approached while 
$N$ increases. 
To be more precise, the BG distribution $p_{BG}$ associated with a 
many-body Hamiltonian ${\cal H}_N$  at thermal equilibrium is given 
by $p_{BG} \propto e^{-\beta {\cal H}_N}$ whenever ${\cal H}_N$ 
includes short-range interactions or no interactions at all. 
We may then write that $p_{BG} \propto e^{-[\beta {\cal H}_N/N]N}$, 
where, consistently with thermodynamics, $[\beta {\cal H}_N/N]$ is 
an intensive quantity. The corresponding LDT statement for a binary stochastic system with 
$N$ random variables yielding $n$ times say 0, and $(N-n)$ times say 1 concerns 
the probability $P_N(n/N > z) \in [0,1]$ of the random variable 
$n/N$ taking values larger than a fixed value $z \in\Re$ for 
increasingly large values of $N$. Under the hypothesis of 
probabilistic independence, or similar settings, we expect 
$P_N(n/N >z) \approx e^{- r_1(z)N}$, where the {\it rate function} 
$r_1$ equals a BG {\it relative entropy per particle}. 
Therefore $r_1(z) N$ plays the role of a 
thermodynamic total entropy which, consistently with the Legendre structure of classical thermodynamics, 
is {\it extensive}, i.e., $r_1(z)N\propto N$ ($N\gg 1)$.
 
Within nonextensive statistical mechanics ({\it $q$-statistics} for short)\cite{Tsallis1988,Santos1997,Abe2000,WilkWlodarczyk2000,Beck2001,BeckCohen2003,HanelThurnerGellMann2014,EncisoTempesta2017,JizbaKorbelLavickaProcksSvdovaBeck2018,JizbaKorbel2019,SuyariMatsuzoeScarfone2020,Tsallisbook}, 
we typically tackle with long-range-interacting Hamiltonian systems, among other strongly correlated ones.
The associated distributions of velocities appear to be 
$Q$-Gaussians with $Q > 1$ 
(see, for instance, \cite{AnteneodoTsallis1998,CirtoAssisTsallis2013,CirtoRodriguezNobreTsallis2018} 
for the $\alpha$-XY ferromagnet, \cite{RodriguezNobreTsallis2019} for the $\alpha$-Heisenberg ferromagnet, 
and \cite{ChristodoulidiTsallisBountis2014,ChristodoulidiBountisTsallisDrossos2016,BagchiTsallis2016,BagchiTsallis2017} 
for the $\alpha$-Fermi-Pasta-Ulam model), with $Q$ approaching unity when the range of the interactions approaches 
the short-range regime. 
These facts are to be associated  with a $Q$-Central Limit Theorem 
($Q$-CLT) which leads, when $N\to\infty$, to a convergence on a 
$Q$-Gaussian distribution. Sufficient conditions for the $Q$-CLT 
to hold are already available \cite{UmarovTsallisSteinberg2008} 
(see also \cite{UmarovTsallisGellMannSteinberg,umarov}) but the 
necessary conditions for a $Q$-CLT still remain as a challenge.
 
Within $q$-statistics we  expect, for the total 
energy of a long-range-interacting system at its stationary, or quasi-stationary, state, to be super-extensive, hence, not proportional to $N$, as it is the case for short-range-interacting systems. More precisely, we expect 
$p_q \propto e_q^{-\beta_q {\cal H}_N}$, where ${\cal H}_N$ is a
super-extensive Hamiltonian, $\beta_q$ playing the role of an inverse {\it effective temperature} ($\beta_1 \equiv \beta$ corresponds to the usual inverse {\it kinetic temperature}); $\beta_q$ generically differs from $\beta$ (in extreme cases, by orders of magnitude \cite{Andradeetal2010}). We remind that $e_q^z \equiv [1+(1-q)z]^{1/(1-q)}$ with $e_1^z=e^z$. For say two-body (attractive) interactions 
decaying like $1/(distance)^\alpha$ ($\alpha \in [0,\infty)$) within a 
$d$-dimensional system, we may rewrite 
$p_q \propto e_q^{-[(\beta_q \tilde{N}) ({\cal H}_N/N\tilde{N})]N}$ 
where $\tilde{N} \equiv \frac{N^{1-\alpha/d}-1}{1-\alpha/d}$ is, for 
$N$ increasingly large, constant for $\alpha/d>1$ (short-range),  
increases like $N^{1-\alpha/d}$ for $0 \le \alpha/d <1$ (long-range), 
and increases like $\ln N$ for $\alpha/d=1$.  

Let us emphasize at this point that both  $ (\beta_q \tilde{N})$ and $({\cal H}_N/N\tilde{N})$ 
are intensive quantities. Indeed, let us illustrate these facts by focusing on the following paradigmatic Gibbs thermodynamical energy:
\begin{eqnarray}
%\begin{split}
&G(V,T,p,\mu,H, \dots) =  U(V,T,p,\mu,H,\dots)  \nonumber \\ 
& - TS(V,T,p,\mu,H,\dots) +pV -\mu N(V,T,p,\mu,H,\dots) \nonumber \\
& -HM(V,T,p,\mu,H,\dots)- \cdots \,,
\label{thermodynamics}
%\end{split}
\end{eqnarray}
where $T, p, \mu,H$ are the temperature, pressure, chemical potential, external magnetic field, and $U,S,V,N,M$ are the internal energy, entropy, volume, number of particles (in turn proportional to the number of degrees of freedom), magnetization. By dividing both sides by $N\tilde{N}$ we obtain
\begin{eqnarray}
%\begin{split}
&\frac{G(V,T,p,\mu,H, \dots)}{N \tilde{N}} =  \frac{U(V,T,p,\mu,H,\dots)}{N \tilde{N}}  \nonumber \\ 
& - \frac{T}{\tilde{N}} \frac{S(V,T,p,\mu,H,\dots)}{N} +\frac{p}{\tilde{N}} \frac{V}{N} - \frac{\mu}{\tilde{N}}  \nonumber \\
& -\frac{H}{\tilde{N}} \frac{M(V,T,p,\mu,H,\dots)}{N}- \cdots \,.
\label{thermodynamics}
%\end{split}
\end{eqnarray}
It has been profusely verified in the literature (see \cite{Tsallisbook,TsallisCirto2013} and references therein) that all the quantities $G/(N \tilde{N})$, $U/(N \tilde{N})$, $T/ \tilde{N}$, $S/N$,  $p/ \tilde{N}$, $V/N$, $\mu/ \tilde{N}$, $H/ \tilde{N}$, $M/N$ are thermodynamically intensive, in the sense that, in the $N\to\infty$ limit, they all yield {\it finite} quantities, thus preserving the Legendre structure of classical thermodynamics for both short- and long-range interactions.

We then identify three classes of thermodynamical variables for all values of $\alpha/d$, namely (i) those that are expected to always be extensive ($S,V, N,M,\ldots$), i.e., scaling with $N$, (ii) those that characterize the external conditions under which the system is placed ($T,p,\mu,H,\ldots$), scaling with $\tilde{N}$, and (iii) those corresponding to energies ($G,U$), scaling with $N \tilde{N}$. For short-range interactions (i.e., $\alpha/d>1$), these three classes collapse into the traditional two (intensive and extensive) currently indicated in the textbooks of thermodynamics.

The desirable mathematical counterpart for such systems
would of course be to have a $q$-Large Deviation Theory 
($q$-LDT) with a probability corresponding to $n/N-1/2 \ge z$ given by $P(N, z)\approx e_q^{- r_q(z)N}$, 
where the {\it rate function} $r_q(z)$ would once again equal some 
{\it relative nonadditive entropy per particle} defined through \cite{Tsallis1988}
\begin{equation}
S_q=k\frac{1-\sum_ip_i^q}{q-1} \;\;(q\in \mathbb{R})\,
\label{qentropy}
\end{equation} 
with $S_1=S_{BG}\equiv -k\sum_ip_i \ln p_i$, $k$ being a conventional positive constant (hereafter taken to be $k=1$). A more precise notation for $r_q(z)$ would be $r_{q_r}(z)$, since, as we shall verify here below, there is no reason for being $q_r=q$; in fact, in all the nontrivial cases that we are aware of, it appears to be  $q_r \ne q$. However, for simplicity, we shall maintain the notation $r_q$.

The quantity $r_q(z) N$ is expected to play a role similar to that of a total system thermodynamic 
entropy which, as mentioned above, should always be extensive, i.e., 
$\propto N$ ($N\gg 1)$. Naturally, in order to unify all the above 
situations, we expect $q=f(Q)$, $f(Q)$ being a smooth function 
which satisfies $f(1)=1$, thus recovering the usual LDT.
 
The above $q$-LDT scenario has already been numerically verified 
for a purely probabilistic model with strong 
correlations \cite{RuizTsallis2012,Touchette2012,RuizTsallis2013}, as well as for diverse 
physical models \cite{TirnakliTsallisAy2021}.

%%%%%%%%%%%%%%%%%%%%%%%%%%%%%%
\section{Model and results}
In the present paper, we focus on  a scale-invariant probabilistic model introduced in \cite{HanelThurnerTsallis2009} 
and based on the Laplace-de Finetti theorem for exchangeable stochastic processes. The random variables are binary 
(Ising-like) and can be either uncorrelated ($Q=1$) or strongly correlated ($Q > 1$); each of them takes the 
values $0$ and $1$. A specific micro-state with $n$ values $0$ and $(N-n)$ values $1$
corresponds to $r_n^N=1/2^N$ 
for $Q=1$ and to
\begin{equation}
r_n^N=\frac{B\Bigl(\frac{3-Q}{2Q-2}+n,\frac{3-Q}{2Q-2} +N-n \Bigr) }{B\Bigl(\frac{3-Q}{2Q-2}, \frac{3-Q}{2Q-2}\Bigr)}
\label{beta}
\end{equation} 
for long-tailed distributions ($1<Q<3$), where $B(x,y)=\Gamma(x)\Gamma(y)/\Gamma(x+y)$ is the Euler Beta function. 
In this model, $(Q-1)$ measures the strength of the global correlations, and varies from zero to 2. 
As a physical analog we may think of a classical $d$-dimensional spin-1/2 highly anisotropic 
XY ferromagnet with two-body interactions decaying as $1/(distance)^\alpha$. This class of systems approach the 
Ising ferromagnet and typically corresponds to magnets with two-body highly anisotropic XY interactions in the 
absence of one-body terms, and also to two-body fully isotropic XY interactions but having, in addition, highly 
anisotropic one-body terms. 
Then $Q=1$ corresponds to $\alpha/d >1$, whereas $Q>1$ is expected to mirror systems with $0<\alpha/d<1$. 
An extreme such case is the $\alpha=0$ one, where all spins interact equally strongly with all the other spins of 
the system. The $Q$-Gaussian distribution would then possibly mirror a non-Maxwellian 
distribution of velocities. Such distributions have been repeatedly observed in similarly complex systems, 
e.g., long-range isotropic XY and Heisenberg ferromagnets as well as the long-range Fermi-Pasta-Ulam $\beta$-model.

In the present model there are $N!/[n!(N-n)!]$ equivalent such micro-states 
($n=0,1,2,\dots,N $). Consistently, we have 
\begin{equation}
\sum_{n=0}^N \frac{N!}{n!(N-n)!} r_n^N=1\,.
\end{equation}

The quantities $r_n^N$ defined in Eq. (\ref{beta}) satisfy (see \cite{HanelThurnerTsallis2009}) the so-called {\it Leibnitz triangle rule}, i.e.,
\begin{equation}
r_n^N +r_{n+1}^N=r_n^{N-1} \;\;(\forall N, \,\forall n)\,.
\label{leibnitz}
\end{equation}
Such sets of probabilities are also known as {\it scale-invariant}  (see also \cite{RodriguezSchwammleTsallis2008,RodriguezTsallis2012,RodriguezTsallis2014}) since their distribution at a given scale $N$ can always be 
obtained by probabilistic marginalization of higher scales (corresponding to $N+1,N+2,\dots$). This property is rather special indeed and should {\it not} be confused with the so-called ``rule of marginalization", which is valid for any distribution of probabilities. An example of well defined distribution of $N$ binary random variables which generically violates relation (\ref{leibnitz}) can be seen in \cite{TsallisGellMannSato2005}. 
The strength of the Leibnitz triangle rule (see, for instance \cite{Polya1945} and related works) can be illustrated by the fact that, under this remarkable hypothesis, the entire set $\{ r_n^N\} \,,\forall (N,n)$, can be generically and univocally recovered by only providing, for instance, the set $\{r_0^N\} \,, \forall N$. In order to provide a more complete view on this issue, let us clarify however that if the system satisfies, as in the present model, exchangeability of the random variables, then the rule of marginalization implies the Leibnitz triangle rule.

The $N\to\infty$ 
attractor of this model for fixed $Q$ turns out \cite{HanelThurnerTsallis2009} to precisely be a $Q$-Gaussian, which makes it an interesting 
case for checking its LDT behavior. 

\begin{figure}[]
\centering
\includegraphics[scale=0.25,angle=0]{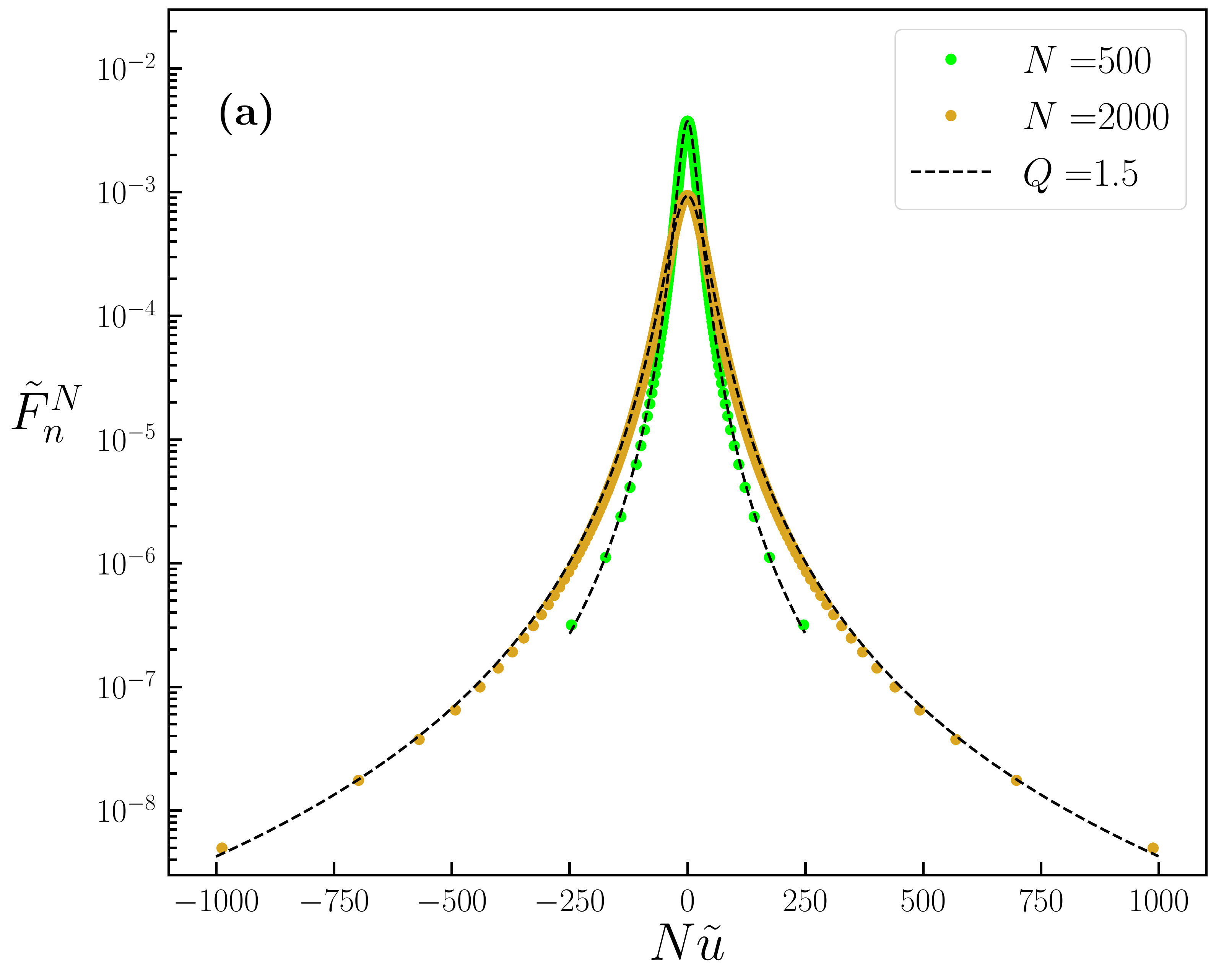}
\includegraphics[scale=0.25,angle=0]{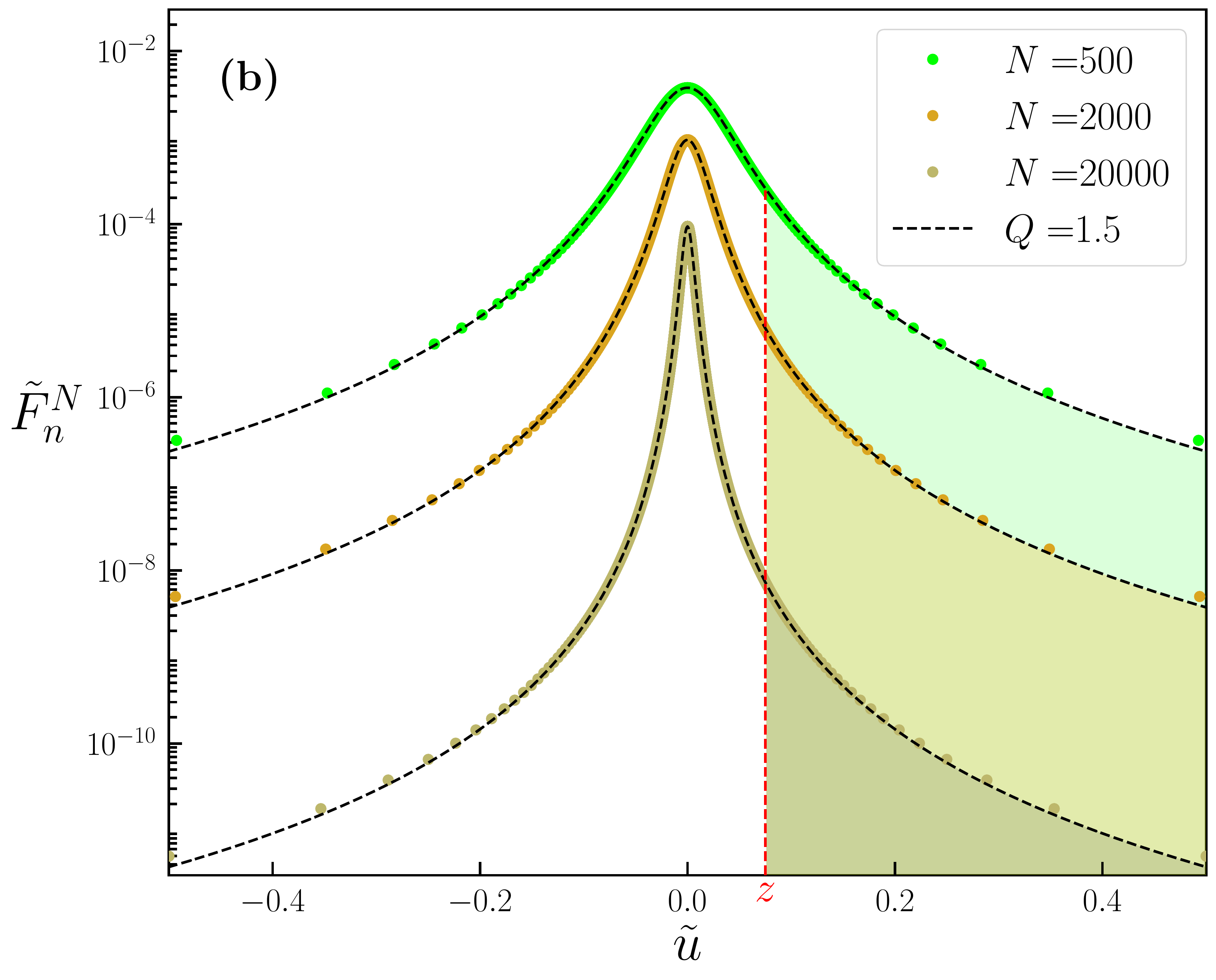}
\caption{\label{QGaussian} {\small 
${\tilde F}_n^N$ distributions are given for some representative values of $N$. 
Dashed lines are the corresponding $Q$-Gaussians ($\propto e_Q^{-2N {\tilde u}^2}$). 
The sums of the values of all points equals unity. Notice a relevant point, namely that the abscissa values 
of these points are not equidistant.
(a) The distribution is represented as a function of $N {\tilde u}\in [-N/2,N/2]$. 
(b) The distribution is represented as a function of $\tilde u \in [-1/2,1/2]$. 
An arbitrary value $z \in [0,1/2]$ is indicated as well. 
}}
\end{figure}

Let us briefly review the algorithm of \cite{HanelThurnerTsallis2009} for constructing the distributions which, 
in the $N\to\infty$ limit, yield exact $Q$-Gaussians. We define
\begin{equation}
u_n^N \equiv \frac {(n/N -1/2)}{ \sqrt{(Q -1)(n/N) (1 - n/N )}}
\end{equation}
and also
\begin{equation}
{\tilde u}_n^N \equiv \frac{u_n^N}{2\,\max_{n=1,2,\dots,N-1}\{u_n^N\}}
\end{equation}
where
$\max_{n=1,2,\dots,N-1}\{u_n^N\}= \frac{\frac{1}{2}-\frac{1}{N}}{\sqrt{(Q-1)(N-1)/N^2}}$. 
We also define the discrete width
\begin{equation}
du_n^N \equiv \frac{[(n/N)(1-n/N)]^{-3/2}}{4(N+1) \sqrt{Q-1}}
\label{width}
\end{equation}
from which it follows the (un-normalized) distribution
\begin{equation}
F_n^N=(du_n^N)^{-1} \frac{N!}{n!(N-n)!)}r_n^N\,,
\end{equation}
where
$r_n^N$ is given by Eq.~(\ref{beta}) and, after normalization, we have 
\begin{equation}
{\tilde F}_n^N \equiv \frac{F_n^N}{\sum_{n=1}^{N-1} F_n^N} \, .
\end{equation}
The values $n=0$ and $n=N$ are excluded from the sum
because these values map to infinity in Eq.~(\ref{width}).
${\tilde F}_n^N$ is the distribution which, for $N\to\infty$, is attracted by a 
$Q$-Gaussian.
In Fig.~\ref{QGaussian}a we have represented the data $\{(N {\tilde u}_n^N, {\tilde F}_n^N)\}$, whereas in 
Fig.~\ref{QGaussian}b the same data have been plotted with respect to ${\tilde u}$ so that the distribution can 
be given in the region $[-1/2,1/2]$. 
Through the rescaling $\bar u \equiv \sqrt{2N} {\tilde u}$, one can easily obtain the normalized attractor 
${\bar F}_n^N(\bar u)=\sqrt{2N}{\tilde F}_n^N(\tilde u)=e_Q^{- {\bar u}^2}$. The discrete width in 
Eq.~(\ref{width}) is related to the nonequidistance observed between the points in Fig.~\ref{QGaussian}. 

Now, in the LDT realm, we focus on the probability $P(N,z)\in[0,1]$ which is defined as that whose values 
of $F_n^N$ correspond to $n/N >1/2+z$. More precisely, it is the sum of all values whose $ \tilde u > z$ 
(see shadowed areas in Fig.~\ref{QGaussian}b). It is clear that $P(N, 0)=1/2$ and $P(N, 1/2)=0$.
In view of what has been discussed above, we expect to  numerically verify that 

\begin{equation}
P(N,z)=P_0(Q,z)\,e_q^{-r_q(Q,z)\,N}\,,
\label{qLDT}
\end{equation}
with $P_0(Q,0)=1/2$, $P_0(Q,1/2)=0$, and $r_q(Q,0)=0$. 

By optimally fitting, with respect to $(q,r_q,P_0)$, the data with Eq.~(\ref{qLDT}), we have heuristically 
found $q$ to be given by the simple composition of the well known additive duality 
($q \to 2-q$) on top of the multiplicative duality ($q \to 1/q$) 
(see, for instance, \cite{TsallisGellMannSato2005,MoyanoTsallisGellMann2006}), namely
\begin{equation}
q=2-\frac{1}{Q} \;\;\;(1 \le Q <3)\,,
\label{conjecture1}
\end{equation}
or, equivalently,
\begin{equation}
\frac{1}{q-1}=\frac{1}{Q-1}+1 \,.
\end{equation}

For uncorrelated binary variables, we have (from \cite{RuizTsallis2012} with $z=x-1/2$) the following 
BG relative entropy (with regard to equal probabilities) 
\begin{eqnarray}
r_1(z)&=&\ln 2 + \frac{1+2z}{2} \ln\frac{1+2z}{2}+\frac{1-2z}{2} \ln \frac{1-2z}{2}    \\
&\sim& 2z^2 + \frac{4}{3}z^4      \;\;\;(z\to 0)\,.
\end{eqnarray}
hence $r_1(z) \in [0,\ln2] $. More generally, for strongly correlated binary variables, we have (from \cite{RuizTsallis2012})
\begin{eqnarray}
r_q(z)&=&\frac{1}{q_r-1} \Bigl\{\frac{1}{2}[(1+2z)^{q_r} +(1-2z)^{q_r}]-1 \Bigr\}    \label{rq}  \\
&\sim& 2q_rz^2 +\frac{2}{3}(3-q_r)(2-q_r)q_rz^4
\;\;(z\to 0),  \label{rqexpanded}
\end{eqnarray}
where we have used $p=1/2+z$ and $1-p=1/2-z$ in expression (\ref{qentropy}), adopting as a reference distribution the equal probability case.
It follows that $r_q(z) \in [0,\ln_{2-q_r}2] $, and we verify that $r_q(z)$ recovers $r_1(z)$ for $q_r \to 1$.

\begin{figure}[]
\centering
\includegraphics[scale=0.25,angle=0]{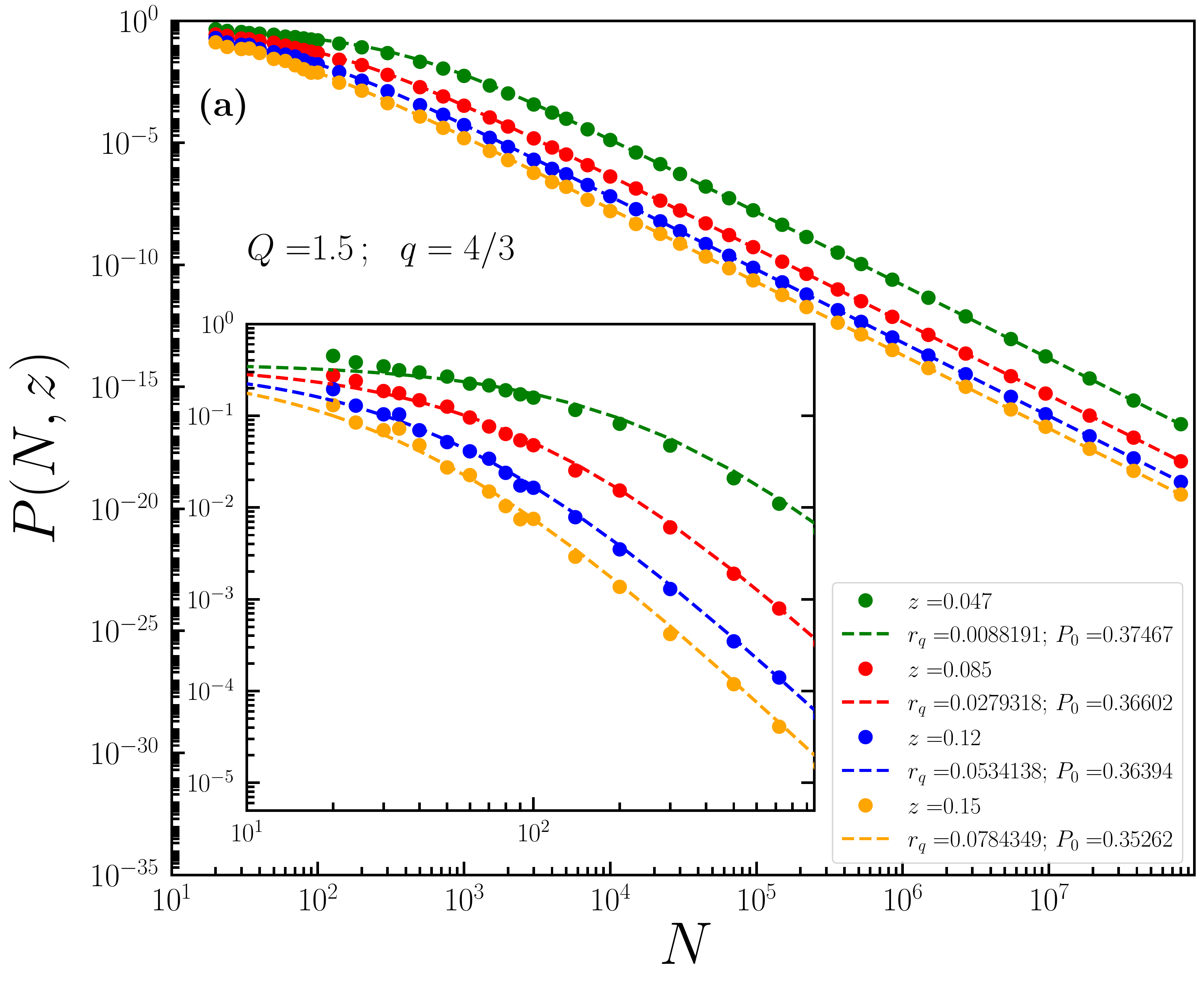}
\includegraphics[scale=0.25,angle=0]{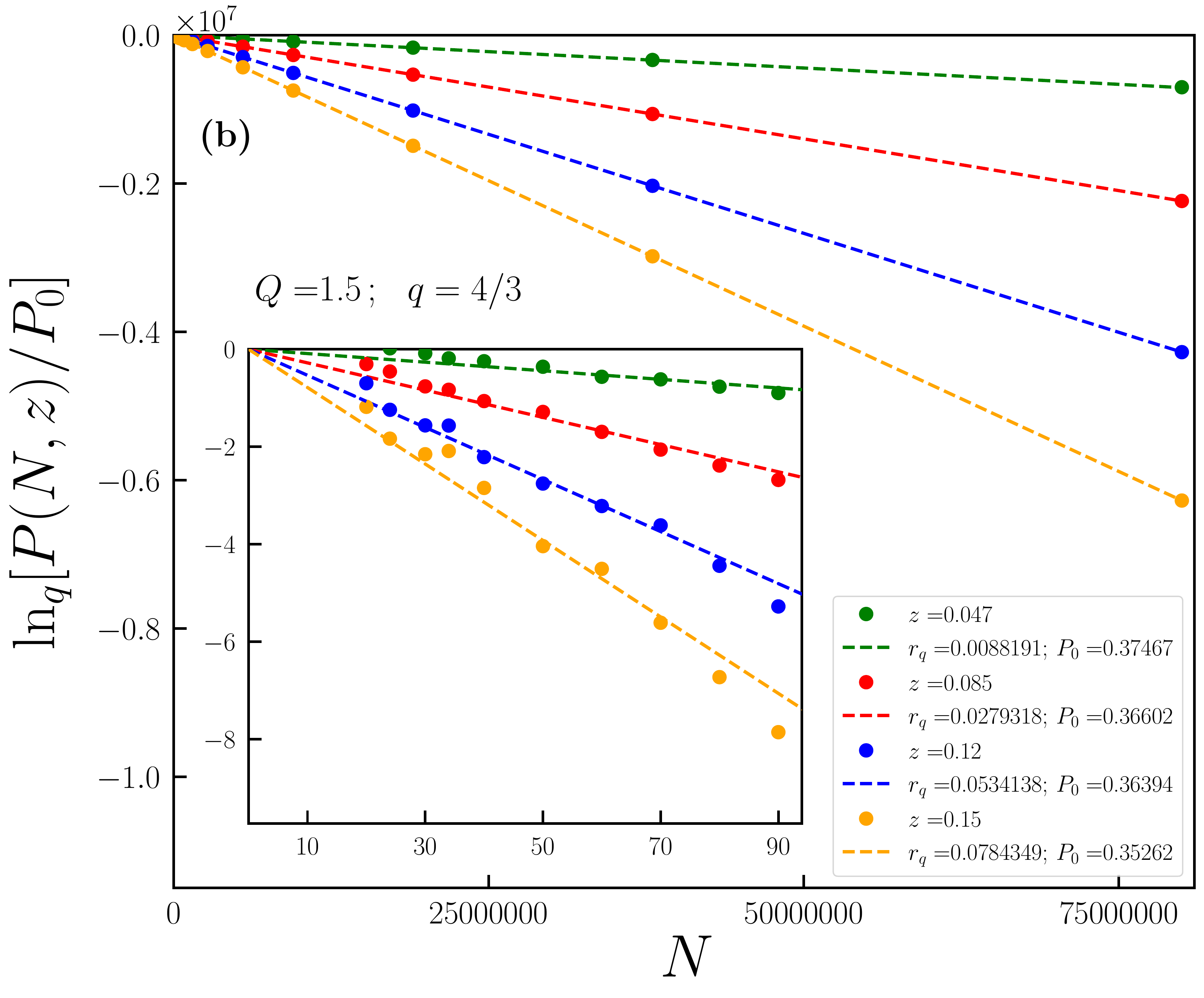}
\includegraphics[scale=0.25,angle=0]{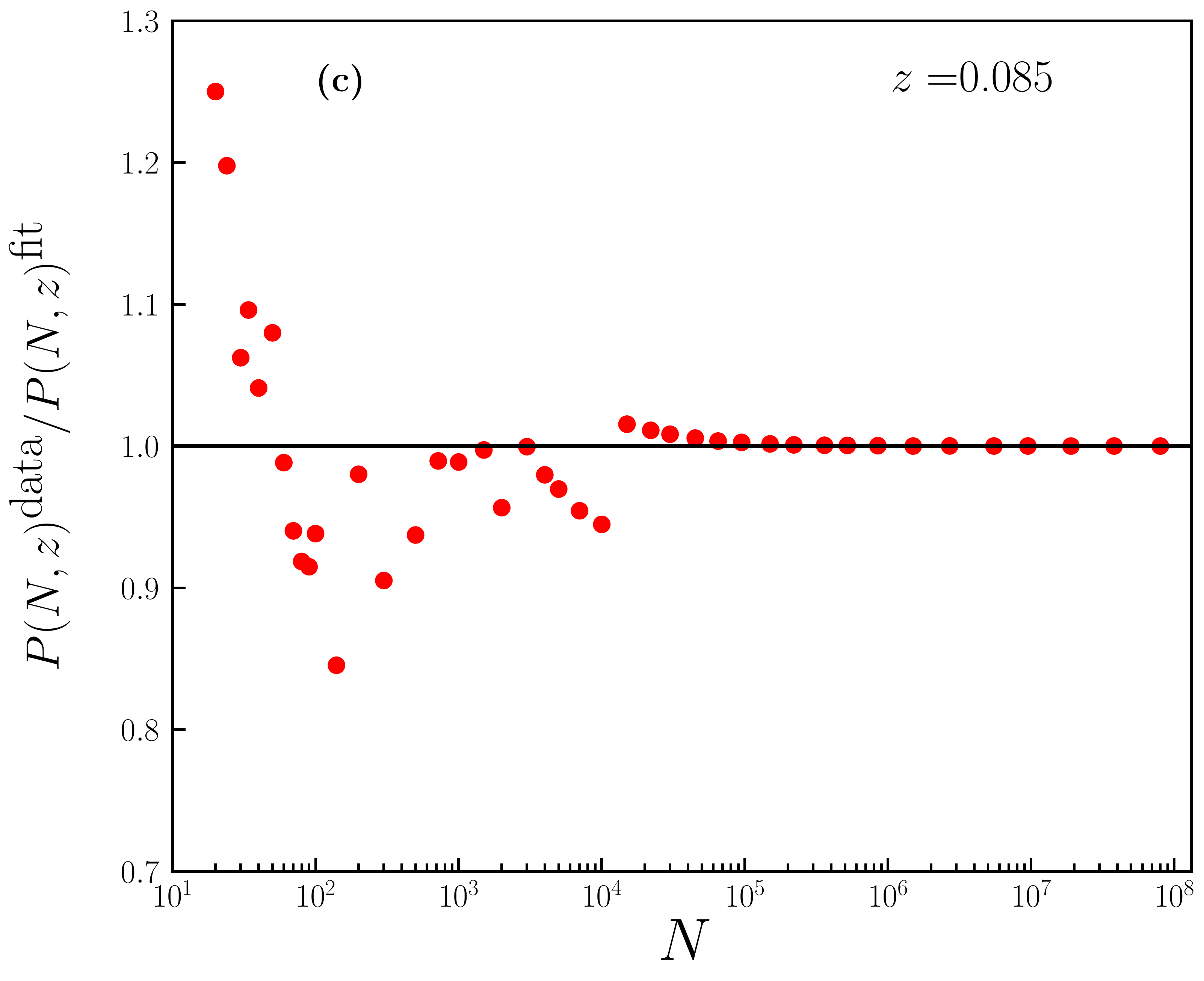}\\
\caption{\label{fig:Fig2} {\small $P(N,z)$ for $Q=1.5$ in log-log (a), $q$-log (b) and ratio (c) 
representations. Note that the only distribution which provides, {\it in all scales}, straight lines 
in a $\ln_q x$ versus $x$ representation is the $q$-exponential function. It is worthy to mention here that, 
in our calculations, $z$ values lies in $[0.035,0.17]$. Calculations for $z$ even larger overcome our 
present computational capacity.
To numerically process the generically chosen $z$ value, we attribute to ${\tilde F}_n^N$ a 5-point cubic 
(polynomial) interpolation using five neighboring data points
(we specifically used interp1d class in scipy.interpolate of Python). 
We have checked for all the values of $z$ but have illustrated in (c) with only one representative example.
}}
\end{figure}

\begin{figure}[]
\centering
\includegraphics[scale=0.25,angle=0]{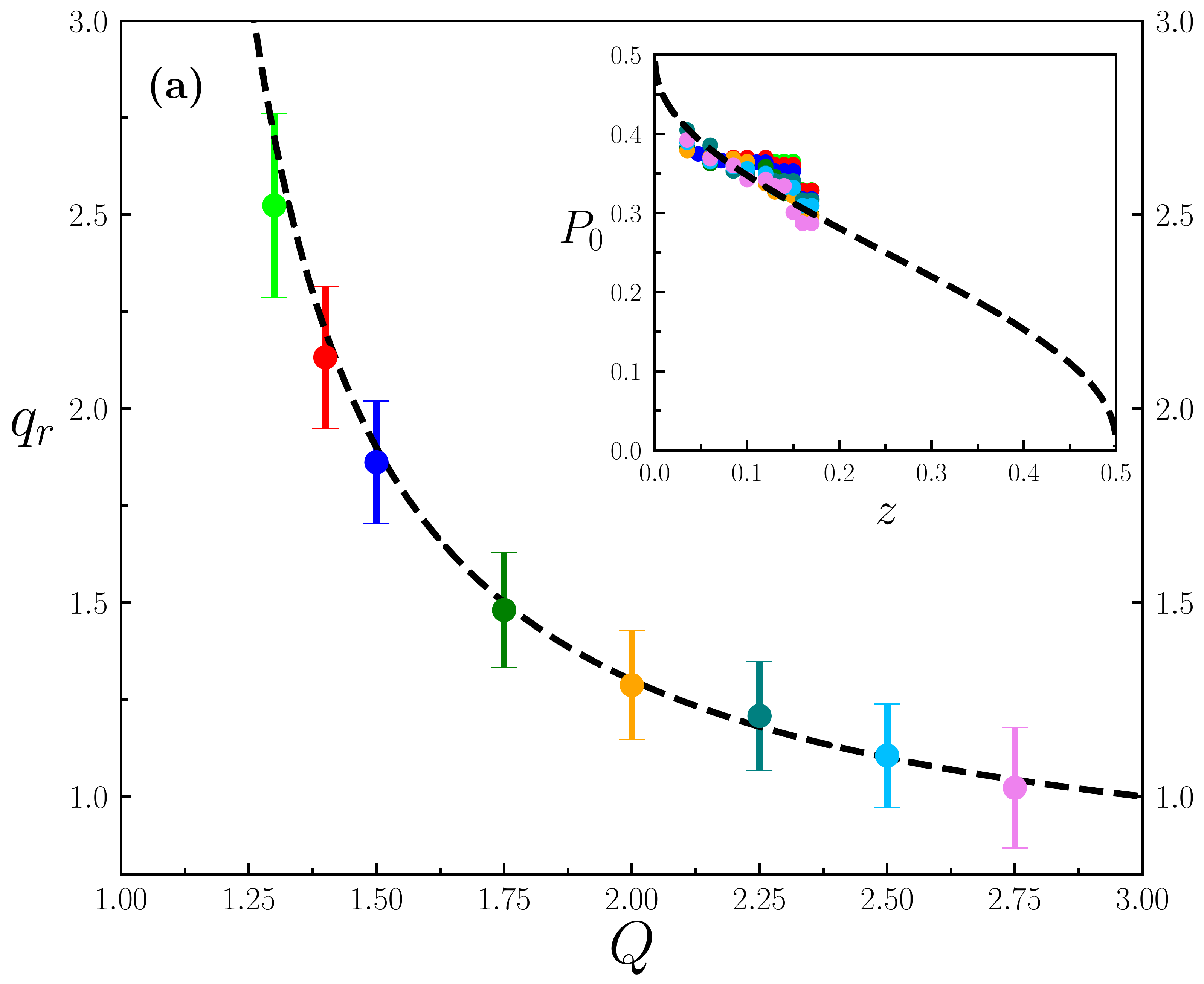}
\includegraphics[scale=0.25,angle=0]{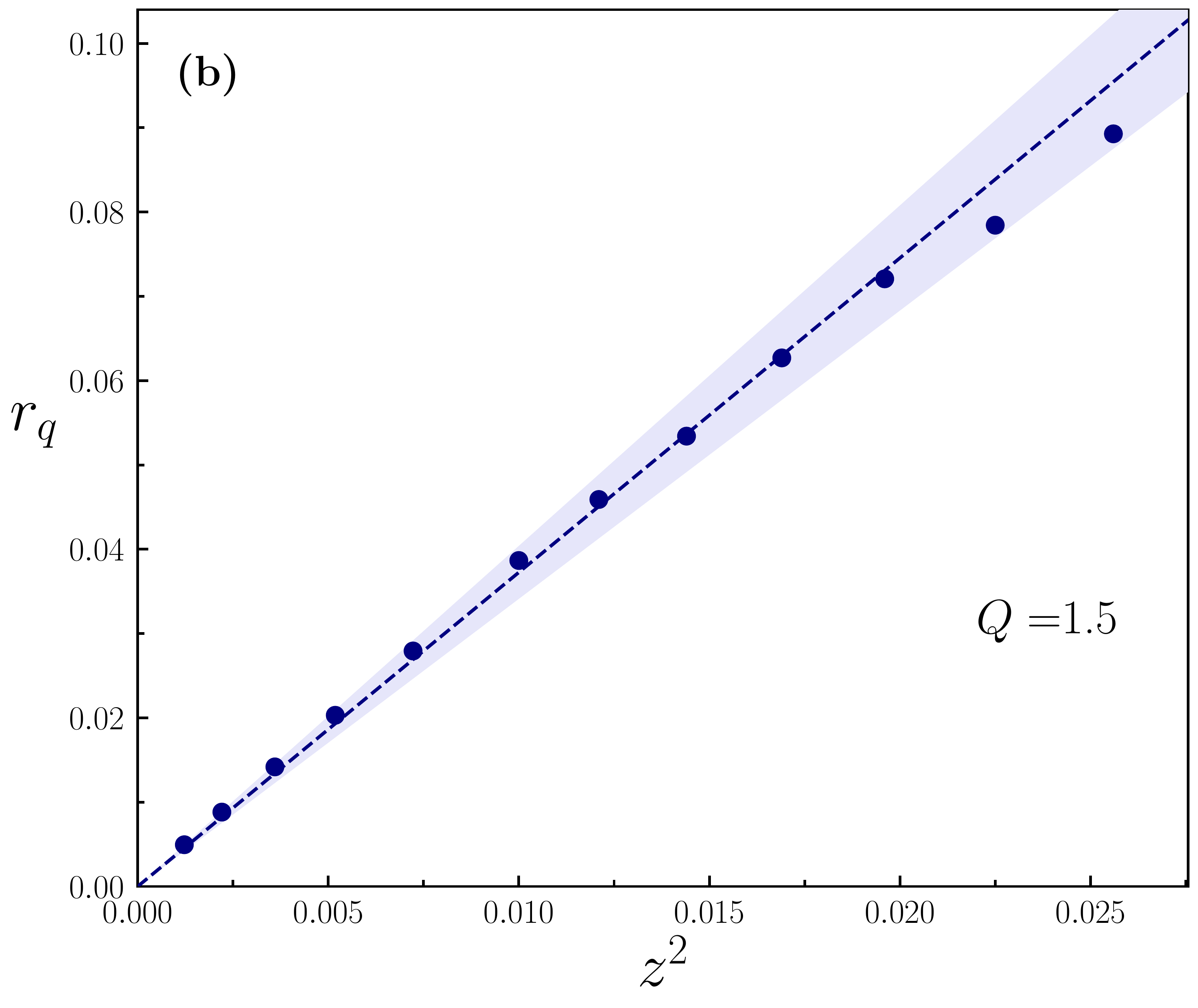}
\caption{\label{fig:Fig3} {\small    (a) $q_r$ versus $Q$ is plotted. 
Dots are obtained through optimization of the overall fitting of $P(N,z)$ with regard to $(P_0 ,q_r)$ 
for typical values of $Q$ and various values for  $z \in [0.035,0.17]$. The optimization procedure uses the 
scipy.optimize module in Python with method SLSQP. The dashed line is the conjecture given in 
Eq.~\ref{conjecture2}. For each value of $z$, we obtain, through an optimized fitting, a value of $q_r$. 
For the whole set of values of $z$, we then obtain the maximal and minimal values of $q_r$, which determine the 
upper and lower error bars. The averaged values are indicated by points.
{\it Inset:} $P_0$ versus $z$ for all $Q$ values seen in the figure; 
the dashed line is given by Eq.~\ref{conjecture3} with the illustrative value $u=0.4$.  
(b) $r_q$ values calculated from Eq.~\ref{rq} are plotted as a function of $z^2$ for a representative $Q$ value. 
The shaded region indicates the interval consistent with the error bars of $q_r$ for this $Q$ value.   
Calculations for $z^2$ even larger overcome our present computational capacity.
}}
\end{figure}

Through the optimized fitting, we heuristically found the following relation
\begin{equation}
q_r = \frac{7}{10} + \frac{6}{10}\frac{1}{Q-1} \;\;\;(1<Q<3).
\label{conjecture2}
\end{equation}
This result was obtained by making $z$ to typically vary up to $0.17$ for diverse values of $Q$, 
which guarantees the verification of the dominant term in Eq.~(\ref{rqexpanded}). If we could numerically 
check up to $z=1/2$, we could guarantee the full expression (\ref{rq}), but this remains out of our present 
computational capacity. 

The numerical determination of $P_0(Q,z)$ is much harder than that of $(q,q_r)$. However, just as a simple 
indication, we have compared the numerical data with 
\begin{equation}
P_0(Q,z) = 1/4 - a z^u + a (1/2 - z)^u \;\;\;(0 \le z \le 1/2))\,,
\label{conjecture3}
\end{equation}
where $a=2^u/4$ in order to satisfy the conditions $P_0(Q,0)=1/2$, $P_0(Q,1/2)=0$. 
This particular form was heuristically proposed as a simple illustration; 
it satisfies $P_0(Q,z)=P_0(Q,1/2-z)$, thus exhibiting an inflexion point at $z=1/4$ 
which was suggested by numerical exploration (see the Inset of Fig. \ref{fig:Fig3} (a)). 
Let us however emphasize here that the precise numerical values of $P_0$ (as well as its unknown exact 
analytical expression) have no particular relevance. They play in fact a rather minor role 
in the conjecture (\ref{qLDT}), in complete similarity with the corresponding pre-factor in the standard LDT. 

As can be seen in Figs.~\ref{fig:Fig2} and \ref{fig:Fig3}, strong numerical evidence supports conjecture 
in Eq.~(\ref{qLDT}) with relations (\ref{conjecture1}), (\ref{rq}), (\ref{conjecture2}) and (\ref{conjecture3}). 
Let us stress that the rate function $r_q$ yields an {\it extensive} (relative) total entropy $r_q N$ for 
all values of $Q\in [1,3)$, as mandated by the Legendre structure of thermodynamics. 
However, the corresponding entropic index is not $q(Q)$ but a different one, namely $q_r(Q)$. 
It remains as an open question whether this is the generic case, or rather an exception associated 
to the present specific model. The fact that three different entropic indices emerge, namely $(Q,q,q_r)$, 
is not particularly surprising, given the Moebius algebra which characterizes $q$-statistics 
(see \cite{TsallisGellMannSato2005}, \cite{GazeauTsallis2019} and references therein).
Another point which deserves emphasis concerns the fact that infinitely many classes of distributions 
exist which asymptotically are power-laws, for $N>>1$. However, only one of these infinite 
functional forms is the $q$-exponential. All these power-laws definitively differ in the non-asymptotic region, 
i.e., for relatively low values of $N$, say $N\sim 50, 100$. This is the reason for which special numerical 
attention has been here devoted to that region of $N$. The evidence that has been achieved strongly points in 
favor of precisely the $q$-exponential form.

%%%%%%%%%%%%%%%%%%%%%%%%%%%
\section{Final remarks}
At this point, let us conclude by reminding that our aim is to approach, within a more general context, 
the fingerprints of Boltzmann-Gibbs statistical mechanics, namely the Maxwellian distribution of velocities 
and the BG exponential weight for the energies. Indeed, in the realm of $q$-statistics based on nonadditive 
entropies, a $Q$-Gaussian distribution emerges for the 
velocities and a $q$-exponential weight emerges for the energies, with $Q \ge q \ge 1$, the equalities 
holding precisely for the BG theory. These generalizations should respectively mirror corresponding generalizations 
of the classical Central Limit Theorem and the Large Deviation Theory. 
This scenario has been successfully verified for a purely probabilistic model (see \cite{RuizTsallis2013} 
and references therein), as well as for some simple physical models \cite{TirnakliTsallisAy2021}. 

Finally, a comparison with previous results might be helpful. In Ref. \cite{RuizTsallis2012}, a completely 
different probabilistic model was focused on. In Ref. \cite{TirnakliTsallisAy2021} we have 
focused on four existing, physically motivated, models.
The model that we focus on here is the one introduced in 
Ref. \cite{HanelThurnerTsallis2009}, where no reference at all exists to a possible $q$-large deviation theory. 
In our present manuscript, it has been possible to numerically discuss (with satisfactory precision 
in some cases) (i) The $(Q,z)$-dependence of the pre-factor $P_0(Q,z)$ (this is the first time such a pre-factor is 
focused on in the literature of complex systems); 
(ii) The possible identification of the rate function $r_q(z)$ with a nonadditive relative 
entropy whose index is $q_r$, definitively different from the index $q$ (a possibility that has never been 
handled before) [a definitive numerical identification of $r_q(z)$ with the $q_r$-entropy for the full range of 
$z$ was not possible because our present computational capacity does not allow us to increase $z$ up to $z=1/2$; 
still, the discussion for $z \simeq 0$ has never been undertaken before]; 
(iii) The $Q$-dependence $q_r(Q)$ as indicated in Eq.~(\ref{conjecture2}), which includes an unexpected 
singularity at $Q=1$ (a feature coming from the specificity of the model introduced in 
Ref. \cite{HanelThurnerTsallis2009}, which, as said above, neatly differs from the model discussed 
in Ref. \cite{RuizTsallis2012}). This $q_r(Q)$ dependence only became accessible due to the mathematical 
fact that all the necessary information is already available in the first asymptotic term, namely in the 
quadratic term of $r_q(z)$ as a function of $z$, as exhibited in Eq. (\ref{rqexpanded}). To the best of our knowledge, 
the  above three points have never before been simultaneously attained for any nontrivial model, and certainly not 
in Refs. \cite{RuizTsallis2012} and \cite{TirnakliTsallisAy2021}. Last but not least, in all four models focused on 
in \cite{TirnakliTsallisAy2021}, the values of $(Q,q)$ are numerical ones, whereas in the present paper we have 
obtained analytical expressions for arbitrary real $Q>1$.

Along the lines of the promising results presented herein, analytical approaches (or very high 
precision numerical approaches) would naturally be very welcome, 
either for the specific models studied here and elsewhere \cite{RuizTsallis2012,TirnakliTsallisAy2021}, 
or in the ambitious form of a $q$-generalized theorem for large deviations based on say a 
$Q$-generalized central limit theorem for an important class of strongly correlated random variables, 
which frequently emerges in physics, geophysics, astrophysics, economics, among other areas. 
We hope that the present numerical indications may stimulate $q$-generalizations of theorems 
such as the G\"artner-Ellis one \cite{Gartner1977,Ellis1984}. In fact, a fascinating research field seemingly 
exists at the crossroad of the large deviation theory \cite{Lanford1973,Ellis1985,Oono1989,Ellis1995,Ellis1999} 
and nonadditive entropies such as $S_q$. Since $S_q$ stands at the foundations of nonextensive statistical 
mechanics, such contributions would naturally enable a deeper understanding of the mathematical structure of 
this current generalization of the Boltzmann-Gibbs theory.

In $q$-statistics, new concepts such as the $q$-triplet appear. This was experimentally verified for the first 
time in \cite{BurlagaVinas2005} and, since then, in a plethora of other systems. No such complex structure 
exists in the Boltzmann-Gibbs theory, where, generically, a unique value for $q$ is admissible, namely $q=1$. 
The various values of the indices $(Q, q(Q), q_r(Q))$ that emerge in the present model, together with one 
more \cite{HanelThurnerTsallis2009} ($q_{entropy}=1$, $\forall Q$, corresponding to the index of the entropic functional 
$S_{q_{entropy}}(N)$  which, for the present exchangeable stochastic system, is thermodynamically extensive, 
i.e., $S_{q_{entropy}}(N) \propto N$ for $N\to\infty$, $\forall Q$) are not yet fully elucidated. 
Some preliminary understanding is available in \cite{TsallisGellMannSato2005,GazeauTsallis2019}, but this 
rich issue still remains as a nontrivial and intriguing open problem.\\

%%%%%%%%%%%%%%%%%%%%%%%%%%%%%%
\section*{Acknowledgments}
The numerical calculations reported in this paper were partially performed at TUBITAK ULAKBIM, 
High Performance and Grid Computing Center (TRUBA resources). 
We acknowledge interesting remarks from E.M.F. Curado and R. Hanel, as well as partial financial support 
from CNPq and Faperj (Brazilian agencies). 
U.T. is a member of the Science Academy, Bilim Akademisi, Turkey and acknowledges partial support from 
TUBITAK (Turkish Agency) under the Research Project number 121F269.

%%%%%%%%%%%%%%%%%%%%%%%%%%%%%%%%


\begin{thebibliography}{99}

\bibitem{statphys1}F. Reif, 
{\it Fundamentals of Statistical and Thermal Physics} 
(Waveland Press, Long Grove, 2008).

\bibitem{statphys2}R.K. Pathria and P.D. Beale, 
{\it Statistical Mechanics} (Academic Press, New York, 2011).

\bibitem{CLT1}P. Billingsley, 
{\it Convergence of Probability Measures} (Wiley, New York, 1968). 

\bibitem{CLT2}N.G. van Kampen, 
{\it Stochastic Processes in Physics and Chemistry} 
(North-Holland, Amsterdam, 1981).

\bibitem{LDT1}R.S. Ellis, 
{\it Entropy, Large Deviations and Statistical Mechanics} 
(Springer, Berlin, 1985).

\bibitem{Peliti}C. Giardina, J. Kurchan and L. Peliti,  
{\it Direct evaluation of large-deviation functions},  
Phys. Rev. Lett. {\bf 96}, 120603 (2006).

\bibitem{LDT2}F. den Hollander, {\it Large Deviations} 
(American Mathematical Society, USA, 2008).

\bibitem{Lacomte1} V. Lecomte and J. Tailleur, 
{\it A numerical approach to large deviations in continuous-time},  
J. Stat. Mech., P03004 (2007).

\bibitem{Touchette2009}H. Touchette,  
{\it The large deviation approach to statistical mechanics},  
Phys. Rep. {\bf 478}, 1-69 (2009).

\bibitem{Lacomte2} T. Nemoto, E. G. Hidalgo and V. Lecomte, 
{\it Finite-time and finite-size scalings in the evaluation of large deviation functions: 
Analytical study using a birth-death process},  
Phys. Rev. E {\bf 95}, 012102 (2017).

\bibitem{Tsallis1988}C. Tsallis, 
{\it Possible generalization of Boltzmann-Gibbs statistics}, 
J. Stat. Phys. {\bf 52}, 479-487 (1988).

\bibitem{Santos1997}R.J.V. dos Santos, {\it Generalization of Shannon's theorem for Tsallis entropy}, 
J. Math. Phys.  {\bf 38}, 4104 (1997).

\bibitem{Abe2000}S. Abe, {\it Axioms and uniqueness theorem for Tsallis entropy}, 
Phys. Lett. A {\bf 271}, 74 (2000).

\bibitem{WilkWlodarczyk2000}G. Wilk and Z. Wlodarczyk, {\it Interpretation of the nonextensivity parameter 
$q$ in some applications of Tsallis statistics and Levy distributions}, Phys. Rev. Lett. {\bf 84}, 2770 (2000).

\bibitem{Beck2001}C. Beck, {\it Dynamical foundations of nonextensive statistical mechanics}, 
Phys. Rev. Lett. {\bf 87}, 180601 (2001).

\bibitem{BeckCohen2003}C. Beck and E.G.D. Cohen, {\it Superstatistics}, Physica A {\bf 322}, 267 (2003).

\bibitem{HanelThurnerGellMann2014}R. Hanel, S. Thurner and M. Gell-Mann,  
{\it How multiplicity determines entropy: derivation of the maximum entropy principle for complex systems}, 
PNAS {\bf 111}, 6905 (2014).

\bibitem{EncisoTempesta2017}A. Enciso and P. Tempesta, {\it Uniqueness and characterization theorems for 
generalized entropies}, J. Stat. Mech. 123101 (2017). 

\bibitem{JizbaKorbelLavickaProcksSvdovaBeck2018}P. Jizba, J. Korbel, H. Lavicka, M. Proks, V. Svoboda and C. Beck, 
{\it Transitions between superstatistical regimes: validity, breakdown and applications}, 
Physica A {\bf 493}, 29 (2018).

\bibitem{JizbaKorbel2019}P. Jizba and J. Korbel, {\it Maximum entropy principle in statistical inference: 
case for non-Shannonian entropies}, Phys. Rev. Lett. {\bf 122}, 120601 (2019).

\bibitem{SuyariMatsuzoeScarfone2020}H. Suyari, H. Matsuzoe and A.M. Scarfone, 
{\it Advantages of q-logarithm representation over q-exponential representation from the sense of scale and 
shift on nonlinear systems}, Eur. Phys. J. Special Topics {\bf 229}, 773-785 (2020). 

\bibitem{Tsallisbook}C. Tsallis, 
{\it Introduction to Nonextensive Statistical Mechanics--Approaching a Complex World} (Springer, New York, 2009). 

\bibitem{AnteneodoTsallis1998}C. Anteneodo and C. Tsallis, 
{\it Breakdown of the exponential sensitivity to the initial conditions: Role of the range of the interaction}, 
Phys. Rev. Lett. {\bf 80}, 5313 (1998).

\bibitem{CirtoAssisTsallis2013}L.J.L. Cirto, V.R.V. Assis and C. Tsallis, 
{\it Influence of the interaction range on the thermostatistics of a classical many-body system}, 
Physica A {\bf 393},  286-296 (2014).

\bibitem{CirtoRodriguezNobreTsallis2018}L.J.L. Cirto, A. Rodriguez, F.D. Nobre and C. Tsallis, 
{\it Validity and failure of the Boltzmann weight}, 
EPL {\bf 123}, 30003 (2018).

\bibitem{RodriguezNobreTsallis2019}A, Rodriguez, F.D. Nobre and C. Tsallis, 
{\it $d$-Dimensional classical Heisenberg model with arbitrarily-ranged interactions: Lyapunov exponents and 
distributions of momenta and energies}, 
Entropy {\bf 21}, 31 (2019).

\bibitem{ChristodoulidiTsallisBountis2014}H. Christodoulidi, C. Tsallis and T. Bountis, 
{\it Fermi-Pasta-Ulam model with long-range interactions: Dynamics and thermostatistics}, 
EPL {\bf 108}, 40006 (2014).

\bibitem{ChristodoulidiBountisTsallisDrossos2016}H. Christodoulidi, T. Bountis, C. Tsallis and L. Drossos, 
{\it Dynamics and Statistics of the Fermi--Pasta--Ulam $\beta$--model with different ranges of particle interactions}, 
JSTAT, 123206 (2016).

\bibitem{BagchiTsallis2016}D. Bagchi and C. Tsallis, 
{\it Sensitivity to initial conditions of $d$-dimensional long-range-interacting quartic Fermi-Pasta-Ulam model: 
Universal scaling}, Phys. Rev. E {\bf 93}, 062213 (2016).

\bibitem{BagchiTsallis2017}D. Bagchi and C. Tsallis, 
{\it Long-ranged Fermi-Pasta-Ulam systems in thermal contact: Crossover from q-statistics to Boltzmann-Gibbs 
statistics}, Phys. Lett. A  {\bf 381}, 1123 (2017).

\bibitem{UmarovTsallisSteinberg2008} S. Umarov, C. Tsallis, S. Steinberg, 
{\it On a $q$-central limit theorem consistent with nonextensive statistical mechanics}, 
Milan J. Math. {\bf 76}, 307 (2008).

\bibitem{UmarovTsallisGellMannSteinberg} S. Umarov, C. Tsallis, 
M. Gell-Mann, S. Steinberg, {\it Generalization of symmetric $\alpha$-stable L\'evy distributions for $q > 1$}, 
J. Math. Phys. {\bf 51}, 033502 (2010).

\bibitem{umarov} M.G. Hahn, X.X. Jiang, S. Umarov, {\it On $q$-Gaussians and exchangeability}, 
J. Phys. A {\bf 43}, 165208 (2010).

\bibitem{Andradeetal2010}J.S. Andrade Jr., G.F.T. da Silva, A.A. Moreira, F.D. Nobre and E.M.F. Curado, 
{\it Thermostatistics of overdamped motion of interacting particles}, Phys. Rev. Lett. {\bf 105}, 260601 (2010).

\bibitem{TsallisCirto2013} C. Tsallis and L.J.L. Cirto,  {\it Black hole thermodynamical entropy}, 
Eur. Phys. J. C {\bf 73}, 2487 (2013).

\bibitem{RuizTsallis2012} G. Ruiz and C. Tsallis, 
{\it Towards a large deviation theory for strongly correlated systems}, Phys. Lett. A {\bf 376}, 2451 (2012).
 
\bibitem{Touchette2012} H. Touchette, 
{\it Comment on ``Towards a large deviation theory for strongly correlated systems"}, 
Phys. Lett. A {\bf 377}, 436 (2013).
 
\bibitem{RuizTsallis2013} G. Ruiz and C. Tsallis, 
{\it Reply to Comment on ``Towards a large deviation theory for strongly correlated systems"}, 
Phys. Lett. A {\bf 377}, 49 (2013).

\bibitem{TirnakliTsallisAy2021}U. Tirnakli, C. Tsallis and N. Ay, 
{\it Approaching a large deviation theory for complex systems}, Nonlinear Dynamics (2021), 
https://doi.org/10.1007/s11071-021-06904-3.

\bibitem{HanelThurnerTsallis2009}R. Hanel, S. Thurner and C. Tsallis, 
{\it Limit distributions of scale-invariant probabilistic models of correlated random variables with the 
$q$-Gaussian as an explicit example}, Eur. Phys. J. B {\bf 72}, 263 (2009).

\bibitem{RodriguezSchwammleTsallis2008}A. Rodriguez, V. Schwammle and C. Tsallis, 
{\it Strictly and asymptotically scale-invariant probabilistic models of $N$ correlated binary random variables 
having {\em q}-Gaussians as $N\to \infty$ limiting distributions}, J. Stat. Mech. P09006 (2008).

\bibitem{RodriguezTsallis2012}A. Rodriguez and C. Tsallis, 
{\it A dimension scale-invariant probabilistic model based on  Leibniz-like pyramids}, 
J. Math. Phys. {\bf 53}, 023302 (2012).

\bibitem{RodriguezTsallis2014}A. Rodriguez and C. Tsallis, 
{\it Connection between Dirichlet distributions and a scale-invariant probabilistic model based on 
Leibniz-like pyramids}, J. Stat. Mech. P12027 (2014).

\bibitem{TsallisGellMannSato2005}C. Tsallis, M. Gell-Mann and Y. Sato, 
{\it Asymptotically scale-invariant occupancy of phase space makes the entropy $S_q$ extensive}, 
Proc. Natl. Acad. Sc. USA {\bf 102}, 15377 (2005).

\bibitem{Polya1945}G. P\'olya, {\it How to Solve It}, (Princeton University Press,1945).

\bibitem{MoyanoTsallisGellMann2006}L.G. Moyano, C. Tsallis and M. Gell-Mann, 
{\it Numerical indications of a $q$-generalised central limit theorem}, 
Europhys. Lett. {\bf 73}, 813 (2006).

\bibitem{GazeauTsallis2019}J.-P. Gazeau and C. Tsallis, 
{\it Mobius transforms, cycles and $q$-triplets in statistical mechanics}, Entropy {\bf 21}, 1155 (2019).

\bibitem{Gartner1977}J. G\"artner, {\it On large deviations from an invariant measure}, 
Teor. Verojatnost. i Primenen. {\bf 22}, 27-42 (1977).

\bibitem{Ellis1984}R.S. Ellis, {\it Large deviations for a general class of random vectors},  
Ann. Probab. {\bf 12}, 1-12 (1984).

\bibitem{Lanford1973}O.E. Lanford, {\it Entropy and equilibrium states in classical statistical mechanics}, 
in {\it Statistical Mechanics and Mathematical Problems}, A. Lenard, Ed., Lecture Notes in Physics 
{\bf 20}, 1-113 (Springer, Berlin, 1973).

\bibitem{Ellis1985}R.S. Ellis, {\it Entropy, Large Deviations, and Statistical Mechanics} (Springer, New York, 1985).

\bibitem{Oono1989}Y. Oono, {\it Large deviation and statistical physics}, Progr. Theoret. Phys. Suppl. 
{\bf 99},165-205 (1989).

\bibitem{Ellis1995} R.S. Ellis, {\it An overview of the theory of large deviations and applications to 
statistical mechanics}, Scand. Actuar. J. {\bf 1}, 97-142 (1995).

\bibitem{Ellis1999}R.S. Ellis, {\it The theory of large deviations: From Boltzmann's 1877 calculation to 
equilibrium macrostates in 2D turbulence}, Physica D {\bf 133} 106-136 (1999).

\bibitem{BurlagaVinas2005}L.F. Burlaga and A.F. Vinas, {\it Triangle for the entropic index $q$ of 
non-extensive statistical mechanic observed by Voyager 1 in the distant heliosphere}, 
Physica A {\bf 356}, 375 (2005).


\end{thebibliography}
\end{document}